\documentclass[12pt]{article}

\usepackage{amssymb,amsmath,exscale}

\usepackage{graphicx}

\usepackage{draftmode}
\usepackage{color}
\definecolor{marin}{rgb}   {0.,   0.3,   0.7} 
\definecolor{rouge}{rgb}   {0.8,   0.,   0.} 
\definecolor{sepia}{rgb}   {0.8,   0.5,   0.} 
\usepackage[colorlinks,citecolor=marin,linkcolor=rouge,
            bookmarksopen,
            bookmarksnumbered
           ]{hyperref}

\newtheorem{lemma}{Lemma}[section]

\newtheorem{theorem}[lemma]{Theorem}

\newtheorem{corollary}[lemma]{Corollary}
\newtheorem{remark}[lemma]{Remark}
\newtheorem{example}[lemma]{Example}

\newtheorem{notation}[lemma]{Notation}
\newtheorem{definition}[lemma]{Definition}
\newtheorem{conclusion}[lemma]{Conclusion}
\numberwithin{equation}{section}
\newcommand{\QED}{\mbox{}\hfill \raisebox{-0.2pt}{\rule{5.6pt}{6pt}\rule{0pt}{0pt}} 
          \medskip\par}             
\newenvironment{Proof}{\noindent
    \parindent=0pt\abovedisplayskip = 0.5\abovedisplayskip
    \belowdisplayskip=\abovedisplayskip{\bfseries Proof. }}{\QED}
\newenvironment{Proofof}[1]{\noindent
    \parindent=0pt\abovedisplayskip = 0.5\abovedisplayskip
    \belowdisplayskip=\abovedisplayskip{\bfseries Proof of #1. }}{\QED}
 
\newenvironment{Remark}{\begin{remark}
                \parindent=0pt \upshape\rmfamily }{\QED\end{remark}}

\newcommand{\dd}{\mathrm{d}}

\newcommand{\jb}{{\boldsymbol{j}}}
\newcommand{\lb}{{\boldsymbol{\ell}}}

\newcommand{\ellb}{{\boldsymbol{\ell}}}
\newcommand{\N}{\mathbb{N}}
\newcommand{\Nc}{\mathcal{N}}

\newcommand{\Pc}{\mathcal{P}}

\newcommand{\R}{\mathbb{R}}

\newcommand{\C}{\mathbb{C}}

\newcommand{\T}{\mathbb{T}}
\newcommand{\Z}{\mathbb{Z}}
\newcommand{\Tc}{\mathcal{T}}
\newcommand{\Lc}{\mathcal{L}}

\newcommand{\Uc}{\mathcal{U}}

\newcommand{\om}{\omega}

\newcommand{\Zc}{\mathcal{Z}}
\newcommand{\ep}{\epsilon}

\newcommand{\Norm}[2]{\|#1\|\left.\vphantom{T_{j_0}^0}\!\!\right._{#2}}         
\newcommand{\SNorm}[2]{|#1|\left.\vphantom{T_{j_0}^0}\!\!\right._{#2}}             
\title{Quasi invariant modified Sobolev norms for semi linear reversible PDEs.}        
\author{Erwan Faou and  Beno\^it Gr\'ebert}       
\begin{document}
\maketitle
\abstract{
We consider a general class of  infinite dimensional reversible differential systems. Assuming a non resonance condition on the linear frequencies, we construct for such systems almost invariant pseudo norms that are closed to Sobolev-like norms. This allows us to prove that if  the Sobolev norm of index $s$ of the initial data $z_0$ is sufficiently small (of order $\epsilon$) then the Sobolev norm of the solution is bounded by $2\epsilon$ during very long time (of order $\epsilon^{-r}$ with $r$ arbitrary). It turns out that this theorem applies to a large class of reversible semi linear PDEs including the non linear Schr\"odinger equation on the d-dimensional torus. We also apply our method to a system of coupled NLS equations which is reversible but not Hamiltonian.

We also notice that for the same class of reversible systems we can prove a Birkhoff normal form theorem that in turn implies the same bounds on the Sobolev norms. Nevertheless the technics that we use to prove the existence of quasi invariant pseudo norms  is much more simple and direct.
}

%newpage

\tableofcontents

%%%%%%%%%%%%%%%%%%%%%%%%%%%%%%%%%%%%%%%%%%%
\section{Introduction}
%%%%%%%%%%%%%%%%%%%%%%%%%%%%%%%%%%%%%%%%%%%

The control of the high index Sobolev norms of the solution of nonlinear partial differential equations during  long time is a difficult and interesting problem, in particular on compact manifolds where there is no dispersion effects and thus no time decay of the solutions of the linear part. Recently a series of works gave a solution to this problem by using the Birkhoff normal form theory applied to some Hamiltonian nonlinear PDEs including  in particular the nonlinear Schr\"odinger equation on a d-dimensional torus and the nonlinear wave equation on the circle (see \cite{Bo96},\cite{Bam03},\cite{BG06},\cite{Greb07} and \cite{Bam07}), the Klein Gordon equation on Zoll manifolds (see \cite{BDGS}) or the nonlinear quantum oscillator on $\R^d$ (see \cite{GIP}). The method consists in  obtaining a normal form for the corresponding Hamiltonian function $H$ in  convenient Sobolev type phase spaces $\Pc_s$ in such a way that in the new variables $H$ decomposes into the sum of a Hamiltonian $N$ (the normal form), whose flow preserves the Sobolev norms, and a remainder Hamiltonian $R$ whose vector field, $X_R$  satisfies (here $r$ is an arbitrary integer and $||\cdot ||_s$ denotes the standart Sobolev norm)
$$
\Norm{X_R(z)}{s}\leq C\Norm{z}{s}^{r+1}, \quad \mbox{for }z\in \Pc_s \mbox{ small enough.}
$$
Then a standard bootstrap procedure shows that if the initial data, $z_0$, is sufficiently small, say $\Norm{z_0}{s}\leq \ep=\ep(r,s)$,  then the solution remains under control, 
$$
\Norm{z(t)}{s}\leq 2\ep\mbox{ during very long time }|t|\leq \frac 1 {\ep^r}.$$
The aim of this paper is to obtain the same dynamical result for reversible PDEs that are not necessarily Hamiltonian by a more direct and simple  method. Actually we generalize to the infinite dimension the classical algorithm of construction of approximate integrals of motion (see for instance \cite{Bam03a} section 4 and references quoted therein). In \cite{Bam03a} this generalization is done for reversible Hamiltonian systems, constructing almost invariant actions but not almost invariant pseudo norms. In this short article we want to stress out that the construction actually works for reversible systems that are not Hamiltonian and that this construction leads directly to bounds on Sobolev norms for a large class of semi linear reversible PDEs. We also mention that our approach is totally self contained.

The reversibility property allows to solve exactly the so called homological equation. Here exactly means that we do not solve it modulo terms in normal form (i.e. corresponding to resonant monomials which are actually absent in the reversible context, see Lemma \ref{lem:homo}).  \\
At the same time the simplification in the resolution of the homological equation has a cost: we can consider only non resonant cases (see Definition \ref{SNR}), whereas the Birkhoff normal form technics (see \cite{BG06} \cite{BDGS} \cite{GIP}) allows to deal with resonant cases. Notice that a similar approach, mimicking for instance  \cite{Greb07}, would provide a Birkhoff normal form result for infinite dimensional reversible system and thus would allow to consider resonant reversible system.

%%%%%%%%%%%%%%%%%%%%%%%%%%%%%%%%%%%%%%%%%%%
\section{Setting of the problem}
%%%%%%%%%%%%%%%%%%%%%%%%%%%%%%%%%%%%%%%%%%%

%%%%%%%%%%%%%
\subsection{Abstract  formalism}
%%%%%%%%%%%%%

We denote $\Nc = \Z^d$ or $\N^d$ (depending on the concrete application) for some $d \geq 1$.  For $a = (a_1,\ldots,a_d) \in \Nc$, we set
$$
|a|^2 = \max\big(1,a_1^2 + \cdots + a_d^2\big). 
$$
We define the set $\Zc = \Nc \times \{ \pm 1\}$. For $j = (a,\delta) \in \Zc$, we define $|j| = |a|$ and we denote by $\overline{j}$ the index $(a,-\delta)$.  By a slight abuse of notation we will also denote by $\Nc$ the set $\{j=(a,+1),\  a\in \Nc\}.$

For $z=(z_j)_{j\in \Zc}\in \C^{\Zc} $ we define the involution $\rho$ via the formula
$$
\rho(z)_j = z_{\bar j}$$
We will say that $z$ is \textit{ real }if $\rho(z)=\bar z$ where $\bar z= (\bar z_j)_{j\in \Zc}$ and where for any $\zeta \in \C$, $\bar \zeta$ denotes the complex conjugate of $\zeta$.

For a given real number $s \geq 0$, we consider the Hilbert space $\Pc_s = \ell_s(\Zc,\C)$ made of elements $z \in \C^{\Zc}$ such that
$$
\Norm{z}{s}^2 := \sum_{j \in \Zc} |j|^{2s} |z_j|^2 < \infty.
$$

Let $\Uc$ be a an open set of $\Pc_s$, let $\omega=(\om_j)_{j\in \Zc} \in \R^\Zc$ and let $F$ be a continuous vector field from $\Uc$ to $\Pc$
$$
F(z)=(F_j(z))_{j\in \Zc}.
$$
We consider the following differential system on $\Pc$
\begin{equation}\label{eq1}
i\dot z_j=\omega_j z_j +F_j(z)\ , \quad j\in \Zc.
\end{equation}

\subsection{Hypothesis}
We first describe the hypothesis needed on the vector field $F$
\begin{itemize}
\item[(H1)] \textbf{Regularity condition}: for all $s>d/2$   the map
$$\left(
\begin{array}{rcll}
\Pc_s &\to& \Pc_s\\[2ex]
z &=& (F_j(z))_{j\in \Zc}
\end{array}
\right)$$
is continuous and has a zero of order at least 2 at the origin, in such a way that 
 $$\Norm{F(z)}{s}\leq C\Norm{z}{s}^2 $$
for $z$ sufficiently small (here $C$ is a constant depending on $s$). 

\item[(H2)] \textbf{Reality condition}:  
$$F_{\bar j}(z)=-\bar F_j(z) \mbox{ for all real }z, \mbox{ i.e. when } \rho(z)=\bar z$$
in such a way that equation $\eqref{eq1}_{\bar j}$ is the complex conjugate of equation equation $\eqref{eq1}_{ j}$ (provided the vector $\omega_j$ satisfies a similar condition, see \eqref{omsym} below) and that the flow $\Phi^t$ associated to the differential system \eqref{eq1} preserves the reality of the initial datum: $\Phi^t(z_0)$ is real for all t when $z_0$ is real. 
\item[(H3)] \textbf{Reversibility condition}: for all $z$, 
$$
\rho(F_j(z))=-F_j(\rho(z)), \quad \forall j\in \Zc \mbox{ and for all real }z 
$$
in such a way that the flow $\Phi^t$ associated with the differential system \eqref{eq1} satisfies
$$\rho(\Phi^t(z))=\Phi^{-t}(\rho(z))$$
for all real $z$.

\end{itemize}

We now translate these hypothesis on the coefficients of the Taylor polynomials of $F$. We first need some more notations:

Let $\ell \geq 3$ be a given integer. For $\jb = (j_1,\ldots,j_r) \in \Zc^r$, we define $\mu(\jb)$ as the third largest integer between $|j_1|,\ldots,|j_r|$. Then we set $S(\jb) = |j_{i_r}| - | j_{i_{r-1}}| $ where $|j_{i_r}|$ and $|j_{i_{r-1}}|$  denote the largest and the second largest integer between $|j_1|,\ldots,|j_r|$. 

In the following, for $\lb= (\ell_1,\ldots, \ell_m)$, we use the notation
$$z_{\lb}= z_{\ell_1}\ldots z_{\ell_m}.$$

\begin{definition}
Let $k \geq 2$, $M > 0$ and $\nu \in [0,+\infty)$, and let
\begin{equation}\label{TM}
F_j(z) = \sum_{m = 2}^k \sum_{\lb \in \Zc^m} a_{j\lb} z_{\lb}, \quad j\in \Zc,
\end{equation}
where $a_{j\ell}$ are complex numbers. 

We say that $F \in \Tc_k^{M,\nu}$ if there exist a constant $C$ depending on $M$ such that 
\begin{equation}
\label{Ereg}
\forall\, m= 2,\ldots,k,\quad \forall\, \lb \in \Zc^m,\quad \forall j\in \Zc, \quad |a_{j\lb}| \leq C \frac{\mu(j,\lb)^{M+\nu}}{\mu(j,\lb)+S(j,\lb)^M}.
\end{equation}

\end{definition}

Notice that this definition is the analog of the polynomial spaces used in \cite{Greb07, Bam07}.
We learn from \cite{Greb07} that if $F\in \Tc^{M,\nu}_k $ then $F$ satisfies the regularity hypothesis (H1) for $s \geq \nu +d/2$ and $\Norm{z}{s} \leq 1$. The best constant in the inequality \eqref{Ereg} defines a norm $\SNorm{Q}{\Tc^{M,\nu}_k}$ for which $\Tc^{M,\nu}_k$ is a Banach space. 
We set
$$
T_k^{\infty,\nu} = \bigcap_{M \in \N} \Tc^{M,\nu}_k
$$
which is a Frechet space.

One easily verifies that  a polynomial vector field $F$ of the form \eqref{TM}  satisfies the reality condition (H2) if and only if 
\begin{equation}
\label{rever}\forall\, m= 2,\ldots,k,\quad \forall\, \lb \in \Zc^m,\quad \forall j\in \Zc, \quad a_{\bar j\bar \lb}=-\bar{a}_{j\lb}
\end{equation} 
and that $F$ satisfies the reversibility condition (H3) if and only if 
\begin{equation}
\label{real}\forall\, m= 2,\ldots,k,\quad \forall\, \lb \in \Zc^m,\quad \forall j\in \Zc, \quad a_{\bar j\bar \lb}=-a_{j\lb}.
\end{equation}
Note that (H2) and (H3) imply that $a_{j\lb} \in \R$. 

\begin{definition}
A vector field  $F$ is in the class $\Tc$ if
\begin{itemize}
\item There exists $s_0 \geq 0$ such that for any $s \geq s_0$, $F\in C(\Uc, \Pc_s)$ for some neighborhood $\Uc$  of the origin in $\Pc_s$. 
\item $F$  exhibits a zero of order at least 2 at the origin. 
\item For all $k \geq 1$, there exists $\nu \geq 0$ such that the Taylor expansion of degree $k$ of $F$ around the origin belongs to $\Tc_k^{\infty,\nu}$. 
\item The coefficients of the Taylor expansion of $F$ satisfy \eqref{rever} and \eqref{real}.
\end{itemize}
\end{definition}

 We now describe the  hypothesis on the frequencies vector.
 
 First we assume the  symmetry 
 \begin{equation}
 \label{omsym}
 \om_{\bar j}=-\om_j, \quad j\in \Zc
 \end{equation}
 which ensures the reversibility of the linear part of \eqref{eq1}.
 We also assume an upper  bound of the frequencies of the form
\begin{equation}
\label{Eboundomega}
\forall\, a \in \Nc, \quad |\omega_a| \leq C |a|^m
\end{equation}
for some constants $C > 0$ and $m > 0$. 

The most important assumption is a non resonances condition which is exactly the same as the condition used in \cite{BG06, Greb07, Bam07}: 
%\begin{definition}\label{SNR} A frequencies vector $\omega \in \R^\Nc$ 
%is \textbf{non resonant} if for
%    any $r\in\N$, there are $ \gamma >0$ and 
%    $\alpha >0$ such that for any $\jb\in \Nc^{r}$ and any $1\leq i\leq 
%    r$, one has 
%    \begin{equation} 
%    \label{A.2} 
%   \Omega_\jb:= \left|\omega_{j_1}+\cdots+\omega_{j_{i}}-\omega_{j_{i+1}}-\cdots 
%    -\omega_{j_{r}} \right|\geq \frac{\gamma}{\mu (j)^{\alpha}}  
%    \end{equation} 
%except if $\{j_{1},\ldots,j_{i}\}=\{j_{i+1},\ldots,j_{r}\}$.    
%\end{definition}   

Let  $\jb = (j_1,\ldots,j_r)\in \Zc^r$, and denote by $j_i = (a_i,\delta_i) \in \Nc \times \{\pm 1\}$ for $i = 1,\ldots,r$. We set 
$$
\Omega(\jb) = 
\delta_1\omega_{a_1} + \cdots  + \delta_r\omega_{a_r}. 
$$
\begin{definition}\label{SNR} A frequencies vector $\omega \in \R^\Zc$ 
is \textbf{non resonant} if for
    any integer $r\geq 3$, there exists two constants $ \gamma >0$ and 
    $\alpha >0$ such that for any $\jb\in \Zc^{i}$ with $1\leq i\leq 
    r$, one has 
    \begin{equation} 
    \label{A.2} 
   |\Omega(\jb) | \geq \frac{\gamma}{\mu (\jb)^{\alpha}}  
    \end{equation} 
except if $\jb = \bar\jb$.    
\end{definition}   
Note that the condition $\jb = \bar\jb$ is equivalent to the fact that $z_\jb$ only depends on the actions,
$I_\ell=z_\ell z_{\bar \ell}$, $\ell \in \Nc$.

%%%%%%%%%%%%%
\subsection{The case of Hamiltonian system}
%%%%%%%%%%%%%
Our setting is very close to the Hamiltonian case. Actually we can endow the phase space $\Pc_s$ with the canonical symplectic structure $i\sum_{j\in \Nc}dz_j\wedge dz_{\bar j}$. Then the linear part of \eqref{eq1} corresponds to the Hamilton equations associated with the harmonic oscillator
$$
H_0= \sum_{j\in \Nc}\om_j z_jz_{\bar j}.
$$
The nonlinear part of \eqref{eq1} is also Hamiltonian if and only if there exists a regular function $P$ such that
\begin{equation}
\label{Ham}
\left\{
\begin{array}{rclll}
F_j&=\displaystyle\frac{\partial P}{\partial z_{\bar j}} & j\in \Nc,\\[2ex]
F_{\bar j}&=-\displaystyle\frac{\partial P}{\partial z_{ j}} & j\in \Nc.
\end{array}
\right.
\end{equation}
In this case the total Hamiltonian function reads
\begin{equation}
\label{Edecomp}
H(z) = H_0(z) + P(z) ,
\end{equation}
and the system \eqref{eq1} can hence be written
\begin{equation}
\label{Eham3}
\left\{
\begin{array}{rcll}
\dot z_j &=& \displaystyle  -i \omega_j z_j - i \frac{\partial P}{\partial z_{\bar j}}(z) & j \in \Nc\\[2ex]
\dot z_{\bar j} &=& \displaystyle i \omega_j z_{\bar j} + i \frac{\partial P}{\partial z_j}(z)& j \in \Nc.  
\end{array}
\right.
\end{equation}
Writing down the Taylor polynomial  of order $k \geq 2$ of the function  $P$ as
$$
P(z) = \sum_{m = 3}^k \sum_{\lb \in \Zc^m} a_{\lb} z_{\lb}, 
$$
the reversibility condition (H2) is equivalent to $P(\rho(z))=P(z)$, which is actually true for $H_0$, while the reality condition equivalent to $P(z)\in \R$ for real $z$ (i.e. $\rho(z)=\bar z$), which again is true for $H_0$.

%%%%%%%%%%%%%%%%%
\section{Statement of the result and applications}
%%%%%%%%%%%%%%%%%

%%%%%%%%%%%
\subsection{Main result}
%%%%%%%%%%%
\begin{theorem}\label{thm:main}
For any $r\geq 3$, there exists $s_0(r)>0$ and for each $s>s_0$ there exist $\epsilon_s>0$, $C_s>0$ and a continuous function 
$$
N_s^{(r)}: B(0, \epsilon_s) \to \R^+
$$
where $B(0, \epsilon_s)$ denotes the ball of radius $\epsilon_s$ centered at the origin in $\Pc_s$, such that
\begin{itemize}

\item[(i)] $\big|N_s^{(r)}(z)-\Norm{z}{s}^2\big|\leq C_s \Norm{z}{s}^3$ 
 for all $z\in B(0, \epsilon_s)$

\item[(ii)] if $t\mapsto z(t)$ is a solution of the reversible system \eqref{eq1} then $$\left| \frac{\dd}{\dd t} N_s^{(r)}(z(t))\right| \leq C_s \Norm{z(t)}{s}^{r+1}$$
for all time $t$ such that $z(t)\in B(0, \epsilon_s)$.
\end{itemize}
\end{theorem}
The proof is postponed to section \ref{sec:proof}.
The dynamical consequences are given in the following

\begin{corollary}\label{cor:main}
For any $r\geq 3$ there exists $s_0(r)>0$ and for each $s>s_0(r)$ there exist $\epsilon_s>0$, $C_s>0$ such that if $z_0\in \Pc_s$ satisties $\Norm{z_0}{s}=\epsilon<\frac{\ep_s}2$ then the solution $z(t)$ of \eqref{eq1} with initial datum $z_0$ is a function in $C^1([-T_\ep,T_\ep], \Pc_s)$ with
$$T_\ep\geq \frac 1 {\epsilon^r}.$$
Furthermore
$$\Norm{z(t)}{s}\leq 2\epsilon, \quad \forall t\in [-T_\ep,T_\ep].$$
\end{corollary}
\begin{Proof} We follow the standard bootstrap argument. Let $t\mapsto z(t)$ be the local solution to \eqref{eq1} with initial datum $z_0$. This solution is defined and of class $C^1$ in an interval $(-T,T)$ for some $T>0$ and we have to prove that $T\geq \frac 1 {\epsilon^r}.$ Take $\ep_s$ given by Theorem  \ref{thm:main} but corresponding to $r+2$ instead of $r$. Let $T_0$ be the supremum of the times $0<t<T$ such that $\Norm{z(t')}{s}\leq 2\epsilon$ for all $t'\in [-t,t].$ As $2\ep<\ep_s$ we can apply assertion (ii) of Theorem \ref{thm:main} to get for $t\in (-T_0, T_0)$,
$$|N_s^{(r)}(z(t))-N_s^{(r)}(z_0)|\leq \left|\int_0^t\frac{\dd}{\dd t} N_s^{(r)}(z(t'))dt'\right|\leq C_s |t| (2\ep)^{r+1}.$$
Then using assertion (i) of the same theorem we deduce that for $t\in (-T_0, T_0)$,
$$
\Norm{z(t)}{s}^2 \leq  \Norm{z_0}{s}^2 + 
 C_s (2\ep)^3 +C_s|t|(2\ep)^{r+3}. 
$$
Therefore, reducing eventually $\ep_s$, we obtain that for $t\in (-T_0, T_0)$ and $|t|\leq \frac 1 {\epsilon^r}$
$$\Norm{z(t)}{s}\leq 3/2\epsilon.$$
Hence by definition of $T_0$ and continuity of $t\mapsto \Norm{z(t)}{s}$ we conclude that $T\geq T_0\geq \frac 1 {\epsilon^r}$.
\end{Proof}

%%%%%%%%%%%%%
\subsection{Examples}
%%%%%%%%%%%%%

%%%%%%%
\subsubsection{Nonlinear Schr\"odinger equation on the torus}
%%%%%%%

We first consider Hamiltonian non linear Schr\"odinger equations of the form
\begin{equation}
\label{E1}
i \partial_t \psi = - \Delta \psi + V \star\psi + \partial_3g(x,\psi,\bar \psi),\quad x \in \T^d
\end{equation}
where $V\in C^\infty(\T^d,\C)$ has real Fourier coefficients, and  $g\in C^\infty(\T^d\times\Uc,\C)$ where $\Uc$ is a neighborhood of the origin in $\C^2$. We assume that for all $z\in \C$, we have $g(x,z,\bar z)=g(x,\bar z, z)$, and that $g(x,z, \bar z) = \mathcal{O}(|zi|^3)$. Notice that for such a semi linear Schr\"odinger equation (i.e. with a nonlinear term that depends only on $x$ and on $\psi(x)$ but not on the derivative of $\psi$), the reality condition, $g(x,z,\bar z)\in \R$,  yields naturally to Hamiltonian equations (i.e. with a nonlinear term that can be written $\partial_3g(x,\psi,\bar \psi))$. In other words, the reversible setting is here more restrictive than the Hamiltonian setting.  The Hamiltonian functional is given by 
$$
H(\psi,\bar\psi) = \int_{\T^d} | \nabla \psi | ^2 + \bar\psi (V \star \psi) + g(x,\psi,\bar\psi) \, \dd x. 
$$

Let $\phi_{a}(x) = \big(\frac{1}{2\pi}\big)^{d/2}\ e^{i a\cdot x}$, $a \in \Z^d$ be the Fourier basis on $L^2(\T^d)$. With the notation $\Nc=\Z^d$ and $\phi_j(x)=\phi_a(\pm x)$ for $j=(a,\pm 1)\in \Zc$ we write
$$
\psi =\sum_{j\in \Nc} z_{j} \phi_{j}(x) \quad \mbox{and}\quad
\bar \psi = \sum_{j\in \Nc} z_{\bar j} \phi_{\bar j}(x).
$$
Further we set
$$
\left\{
\begin{array}{rclll}
F_j&=\displaystyle\frac{\partial P}{\partial z_{\bar j}} & j\in \Nc,\\[2ex]
F_{\bar j}&=-\displaystyle\frac{\partial P}{\partial z_{ j}}&j\in \Nc,
\end{array}
\right.
$$
where 
$$
P(z)= \int_{\T^d}g(x,\sum_{j\in \Nc} z_{j} \phi_{j}(x),\sum_{j\in \Nc} z_{\bar j} \phi_{\bar j}(x))\dd x.$$
Then equation \eqref{E1} can (formally) be written
\begin{equation}
i\dot z_j= \om_j z_j + F_j(z), \quad j\in \Zc
\end{equation}
where the frequency vector $\omega_j$ defined by $\omega_{a} =|a|^2 +\hat V_a$ for $a \in \Nc$ and the relation $\omega_{j} = - \omega_{\bar j}$ for all $j \in \Zc$, satisfies  \eqref{Eboundomega} with $m = 2$. 
Now the hypothesis\footnote{This hypothesis is for instance satisfied when $g$ only depends on the modulus of $|\psi|^2$ like in the Gross-Pitaevskii equation. Notice that this condition is not necessary in the Hamiltonian case.}
$$
g(x,\psi,\bar\psi) = g(x,\bar\psi,\psi),
$$
 ensures that $P(\rho(z))=P(z)$ and thus implies the reversibility condition (H2). The reality condition is also satisfied since $g(x,z,\bar z)$ is real.
The fact that the nonlinearity 
$F$ belongs to $\Tc$ can be verified using the regularity of $g$ and the properties of the basis functions $\phi_a$, see \cite{Greb07,BG06}. In this situation, it can be shown  that the non resonance condition is fulfilled for a large set of potential $V$ (see \cite{BG06} or \cite{Greb07}). 
%%%%%%%%%%%%%%%%%%%%%%%%%%%%%%%%%%
\subsubsection{Coupled NLS on the torus}
\label{coupledNLS}
%%%%%%%%%%%%%%%%%%%%%%%%%%%%%
To generate a reversible PDE that is not Hamiltonian, we have to consider systems of coupled PDEs. 
As example of a system of coupled partial differential equations we
consider a pair of NLS equations coupled via the nonlinear terms. This kind of system is used in nonlinear optics (see for instance \cite{Agr-Boy92,New-Mol92} and references quoted therein). 
From the mathematical point of view the interest of this example is
that the reversible context is much more rich than the Hamiltonian one.
We consider the system for $(\psi,\phi)$ given by
\begin{eqnarray}
\label{cou.2}
i\dot \psi&=& -\psi_{xx}+V_1\star \psi+\partial_{\bar \psi}g_1(x,\psi,\bar \psi, \phi,\bar \phi),
\\
\label{cou.3}
i\dot \phi&=& -\phi_{xx}+V_2\star \phi+\partial_{\bar \phi}g_2(x,\psi,\bar \psi, \phi,\bar \phi). 
\end{eqnarray}
Assume as in the previous section that $V_1, \ V_2\in C^\infty(\T^d,\C)$ have real Fourier coefficients,  and that $g_1,\ g_2\in C^\infty(\T^d\times\Uc,\C)$ where $\Uc$ is a neighborhood of the origin in $\C^2$. We assume that $g_i(x,z,\bar z,\zeta, \bar \zeta)=g_i(x,\bar z, z,\bar \zeta, \zeta)$, and that $g_i(x,z, \bar z,\zeta, \bar \zeta) = \mathcal{O}(|z|^3+|\zeta|^3)$ for $i=1,2$. Thus the system fulfills the three conditions : reality, reversibility and regularity. Nevertheless, in general this system is Hamiltonian  only if $g_1=g_2$. For instance take $g_1(\psi,\bar \psi, \phi,\bar \phi)=|\psi|^4 |\phi|^2 $ and $g_2(\psi,\bar \psi, \phi,\bar \phi)=|\psi|^2|\phi|^2$ to obtain a reversible but non Hamiltonian system.
 
The frequencies are given by
$$
\omega^1_{a}:=|a|^2+\hat V_1(a)\ ,\quad \omega^2_{a}:=|a|^2+\hat V_2(a)\ ,\quad a\in \Z^d.
$$
We can adapt results of  \cite{Greb07} to prove that  the non resonances condition   is fulfilled for a large set of potential $(V_1,V_2)$ (see also \cite{BG06} section 3.4).

%%%%%%%%%%%%%%%%%
\section{Proof of the main theorem}\label{sec:proof}
%%%%%%%%%%%%%%%%%
We adapt the classical algorithm of construction of the approximate integrals of motion (see for instance \cite{Bam03a,FC08}). We Taylor expand the vector field $F$ as 
\begin{equation}\label{F}
F=\sum_{k=2}^{r}F^{(k)} + \tilde F^{(r+1)}
\end{equation}
where 
\begin{equation}\label{tildeF}
F^{(k)}= (F^{(k)}_j)_{j\in \Zc}\in \Tc^{\infty,\nu}_k. 
\end{equation}
Here, each $F^{(k)}_j$ is a homogeneous polynomial of degree $k$ and $ \tilde F^{(r+1)}$ is a remainder term satisfying
$$
\Norm{ \tilde F^{(r+1)}(z)}{s}\leq C_s \Norm{z}{s}^{r+1}.
$$
for some constant $C_s$ depending on $s$. 
We search the almost invariant pseudo norm  $N_s^{(r)}$ under the same form:
\begin{equation}\label{N}
N_s^{(r)}(z)= \sum_{k=2}^{r}N_{s,k}(z)
\end{equation}
where for all $k \geq 2$ $N_{s,k}(z)$ is an even (i.e. satisfying $N_{s,k}(\rho(z)) = N_{s,k}(z)$ for all $z$) continuous homogeneous polynomial of degree $k$  with in particular 
$$
N_{s,2}(z)=\Norm{z}{s}^2.
$$ 
More precisely we will search the polynomials $N_{s,k}(z)$ in the class $\Gamma^\gamma_k$ that we now define: 
\begin{definition} Let $\gamma>0$ and $k \in \N$.
A formal homogeneous  polynomial of degree $k$ on $\Pc_s$ 
$$Q(z)=\sum_{\jb \in \Zc^k}b_\jb z_\jb$$
is in the class $\Gamma^\gamma_k$ if there exists a constant $C>0$ such that 
$$
|b_\jb|\leq C \mu(\jb)^\gamma \frac{\beta(\jb)^s}{(1+S(\jb))^2}
$$
for all ${\jb \in \Zc^k}$ where $\mu(\jb)$ and $ S(\jb)$ are defined in section 2 and $\beta(\jb)$ is the product of the two largest index between $|j_1|,\ldots,|j_k|$.\\
Furthermore we say that $Q$ is even (resp. odd) when
$$b_\jb =b_{\bar \jb}, \quad (\mbox{resp.} \quad b_\jb =-b_{\bar \jb}) \quad  {\jb \in \Zc^k}.$$
\end{definition}
Notice that $z\mapsto \Norm{z}{s}^2$ is in $\Gamma^0_2$ (in that case we have always $j_1=j_2$ and by convention $\mu(\jb)=1$). Remark also  that even polynomials  are real valued for real $z$, provided the coefficients $b_\jb$ are all real. 
The proof of the following lemma is postponed to the Appendix
\begin{lemma}\label{lem:1}

(i) If $0\leq\gamma< s-1/2$, then for all $k \geq 3$ the space $\Gamma^\gamma_k$ is included in the space of continuous polynomials from $\Pc_s$ to $\R$ and in particular if $Q\in \Gamma^\gamma_k$, there exists a constant $C>0$ (depending on $s$ and $k$) such that
$$
|Q(z)|\leq C\Norm{z}{s}^k.
$$
(ii) Let $Q\in \Gamma^\gamma_k$ with $0\leq \gamma < s-1/2$ and $k \geq 3$, then the map $z\mapsto \nabla Q$ is continuous from $\Pc_s$ into $\Pc_{-s}$ and in particular, if $Q\in \Gamma^\gamma_k$, there exists a constant $C>0$ such that
$$
\Norm{\nabla Q(z)}{-s}\leq C\Norm{z}{s}^{k-1}.
$$
\end{lemma}
The second technical lemma whose proof is again postponed to the appendix linkgs the two classes of polynomials defined above. For a vector field $F$ and functional $G$ defined on $\Pc_s$ let us define (formally) the Lie derivative of $G$  along $F$ by
$$\Lc_F G:= \sum_{\ell \in \Zc} F_\ell \frac{\partial G}{\partial z_\ell}.
$$
Then the following result holds true:
\begin{lemma}\label{lem:2}
Let $\gamma, \nu \geq 0$ be given real numbers and $m, n \geq 2$ two integers. For a given $s$,
let $G\in \Gamma^\gamma_n$ and $F\in \Tc^{\infty,\nu}_m$, and assume $s\geq \max(\gamma, \nu +4)$ then 
$\Lc_F G\ \in \Gamma^{\gamma+\nu +4}_{n+m-1}.$ Moreover if $F$ satisfies the Hypothesis (H2)  and (H3), then if $G$ is even $\Lc_F G$ is odd. 
\end{lemma}
For $Q\in \Gamma^\gamma_k$ for some $\gamma > 0$ and $k \geq 2$ we define the $\omega$-derivative $\partial_\omega$ by the formula
$$
\partial_\omega Q(z)=\sum_{\ell \in \Zc} \om_\ell z_\ell \frac{\partial Q}{\partial z_\ell}.
$$
The key to prove Theorem \ref{thm:main} relies on the construction of iterative solutions of homological equations. The next Lemma shows how it is possible to solve them. 
\begin{lemma}\label{lem:homo}
Let $k$ be an integer, let $G\in \Gamma^\gamma_k$ be an odd homogeneous polynomial of degree $k$ and let $\om$ be a non resonant vector of frequencies satisfying \eqref{A.2}.
The homological equation
$$
\partial_\om N= G$$
has a unique solution $N\in  \Gamma^{\gamma+\alpha}_{k}$ which is furthermore an even polynomial.
\end{lemma}
\begin{Proof}  Write 
$$G(z)=\sum_{\jb \in \Zc^k}a_\jb z_\jb$$
and search 
$$N(z)=\sum_{\jb \in \Zc^k}b_\jb z_\jb$$
satisfying the homological equation $\partial_\om N= G$.  With these notations, the last equation is equivalent  to
\begin{equation}\label{triv}
\Omega(\jb)b_\jb=a_\jb, \quad \jb \in \Zc^k.
\end{equation}
Notice that, since $G$ is odd, $a_\ellb=-a_{\bar \ellb}$ and thus $a_\ell=0$ when $\ellb=\bar \ellb$. But, as $\omega$ is non resonant, 
$$\Omega(\jb)=0 \iff \jb= \bar \jb.$$
Therefore equation \eqref{triv} is always solvable by setting
$$b_\jb=\Omega(\jb)^{-1}a_\jb, \quad \jb \in \Zc^k.$$
Then, the fact that $N$ belongs to $ \Gamma^{\gamma+\alpha}_{k}$ is a consequence of \eqref{A.2}. Furthermore, since $\Omega(\bar \jb)=-\Omega(\jb)$ and $G$ is odd, we deduce that $N$ is even.
\end{Proof}

\begin{Remark}
In the Hamiltonian case, this miracle does not occur: the homological equation cannot be solve exactly and we have to add so called normal terms which correspond to the resonant monomials $ z_\jb$ with  $\jb= \bar \jb$. (see for instance \cite{Greb07})\end{Remark}
\begin{Remark}
In the previous Lemma, if the coefficients of $G$ are real, then as the frequencies $\omega_j$ are real, the coefficients of $N$ remain real. 
\end{Remark}
Now we have the tools to prove Theorem \ref{thm:main}.

\begin{Proofof}{Theorem \ref{thm:main}}

We have
$$
\frac{\dd}{\dd t} N_s^{(r)}=-i\sum_{j \in \Zc}\left(\om_j z_j+F_j(z)\right) \frac{\partial N_s^{(r)}}{\partial z_j}
$$
Inserting the Taylor expansions  \eqref{F} and \eqref{N} and equating the terms of the same degree we get the recursive homological equations, $k=2,\ldots,r-1$,
$$
\partial_\om N_{s,k+1}=G_{k+1}
$$
where $G_k$ is determined by
\begin{equation}
\label{eq:G}
G_{k+1} = - \sum_{m=2}^{k}\Lc_{F^{(k+2-m)}} N_{s,m},\quad k=2,\ldots,r-1 .  
\end{equation}
Now by Lemma \ref{lem:2} and Lemma \ref{lem:homo}  these formal equations can be solved verifying at each step that $G_{k}\in \Gamma_{k}^{(k-3)\alpha}$ is odd and that  $N_{s,k}\in \Gamma_{k}^{(k-2)\alpha}$ is even. Moreover as the coefficients of the vector fields $F^{(m)}$, $m = 2,\ldots,r$ are real, we see that the coefficients of the polynomials $N_{s,k}$ remain real at each step.

We then verify estimate (i)  by using Lemma \ref{lem:1}. To verify (ii) we remark that by construction
$$
\frac{\dd}{\dd t} N_s^{(r)}=Q_{r+1}-i\sum_{j \in \Zc}\tilde F_j \frac{\partial N_s^{(r)}}{\partial z_j}
$$
where $Q_{r+1}$ is a polynomial of degree $r+1$ in $\Gamma_{r+1}^{(r-2)\alpha}$. Thus using again Lemma \ref{lem:2} we have $\Norm{Q_{r+1}(z)}{s}\leq C\Norm{z}{s}^{r+1}$. On the other hand
$$
\left|\sum_{j \in \Zc}\tilde F_j(z) \frac{\partial N_s^{(r)}}{\partial z_j}(z)\right| \leq
\Norm{\tilde F(z)}{s}\Norm{\nabla N_s^{(r)}(z)}{-s}$$
and we conclude using \eqref{tildeF} and Lemma \ref{lem:1} that
$$\left|\sum_{j \in \Zc}\tilde F_j(z) \frac{\partial N_s^{(r)}}{\partial z_j}(z)\right| \leq C\Norm{z}{s}^{r+1}.$$

\end{Proofof}

%%%%%%%%%%%
%\section{Extension to a Birkhoff Normal form result for reversible systems}
%%%%%%%%%%%

%%%%%%%%%%%%
\section{Appendix: Proof of the two technical lemmas}
\begin{Proofof}{Lemma \ref{lem:1}}

(i) Let $Q(z)=\sum_{\jb \in \Zc^k}b_\jb z_\jb$ be in $\Gamma^\gamma_k$ with $\gamma<s-1/2$. We have
\begin{align*}
|Q(z)|&\leq C\sum_{\jb \in \Zc^k} \mu(\jb)^\gamma \frac{\beta(\jb)^s}{(1+S(\jb))^2} |z_\jb|\\
&\leq C \sum_{\jb \in \Zc^k} \mu(\jb)^\gamma \frac{\beta(\jb)^s}{(1+S(\jb))^2 \Pi_{m=1}^{k}|j_m|^{s}} \Pi_{m=1}^{k}|j_m|^{s}|z_{j_m}|,
\end{align*}
where we used the notation $\jb = (j_1, \ldots, j_k)$ for a generic multi-index in $\Zc^k$. 

By symmetry of the right hand side of the last inequality, we can reduce the sum to the indices $\jb$ that are ordered in the sense that $|j_1|\geq |j_2|\geq \ldots \geq |j_k|$ so that $\beta(\jb) = |j_1||j_2|$.

First remark that by Cauchy-Schwarz inequality, for any $s_0>1/2$, one has
\begin{equation}
\label{peou}
\sum_\ell |\ell|^{s-s_0}|z_\ell|\leq C \Norm{z}{s}
\end{equation}

Then we obtain using Cauchy-Schwarz inequality for each index $j_4,\ldots, j_k$
\begin{align*}
|Q(z)|&\leq C\sum_{\jb \in \Zc^k} \frac 1 {(1+S(\jb))^2|j_3|^{s-\gamma} \Pi_{m=4}^{k}|j_m|^{s}} \Pi_{m=1}^{k}|j_m|^{s}|z_{j_m}| \\
&\leq C\Norm{z}{s}^{(k-3)} \sum_{|j_1|\geq |j_2|\geq  |j_3|} \frac 1 {(1+|j_1|-|j_2|)^2|j_3|^{s-\gamma} }\Pi_{m=1}^{3}|j_m|^{s}|z_{j_m}| .
\end{align*}
Now as  $s-\gamma>1/2$, and using again the Cauchy-Schwarz inequality and \eqref{peou}, we have that
$$
\sum_{j_3}\frac{|j_3|^{s}|z_{j_3}|}{|j_3|^{s-\gamma}}\leq C\Norm{z} {s}.
$$
Eventually, as $j\mapsto(1+|j|)^{-2}$ is a $\ell^1$ sequence and as the convolution product of a $\ell^1$ sequence with a $\ell^2$ sequence gives rise to a $\ell^2$ sequence, we obtain
$$
\sum_{|j_1|\geq|j_2|} \frac 1 {(1+|j_1|-|j_2|)^2}|j_2|^{s}|z_{j_1}||j_2|^{s}|z_{j_2}|\leq  C\Norm{z}{s}^2$$
which concludes the proof of assertion (i).

\medskip

\noindent (ii) Let $Q(z)=\sum_{\jb \in \Zc^{k+1}}b_\jb z_\jb$ be in $\Gamma^\gamma_k$ with $\gamma<s-1/2$. We have
\begin{align*}
\Norm{\nabla Q(z)}{-s}^2&=\sum_{\ell \in \Zc} |\ell|^{-2s}\left| \frac{\partial Q}{\partial z_\ell}\right|^2\\
&\leq C \sum_{\ell \in \Zc} |\ell|^{-2s}\left(\sum_{\jb \in \Zc^{k}}|b_{\ell,\jb}| |z_{j_1}|\ldots|z_{j_k}|\right)^2\\
&\leq C(k!)^2\sum_{\ell \in \Zc} |\ell|^{-2s}\left(\sum_{\jb \in \Zc_>^{k}}|b_{\ell,\jb}| |z_{j_1}|\ldots|z_{j_k}|\right)^2
\end{align*}
where $\Zc_>^k$ denotes the set of ordered $k$-uples $(j_1,\ldots,j_k)$ such that $|j_1|\geq \ldots\geq |j_k|$. 
Then
we use again \eqref{peou} to obtain
\begin{multline*}
\Norm{\nabla Q(z)}{-s}^2 
\leq C\Norm{z}{s}^{2(k-3)} \\[2ex] \times \sum_{\ell \in \Zc} |\ell|^{-2s}\left(\sum_{|j_1|\geq|j_2|\geq|j_3|} \frac{\mu(\ell,j_1,j_2,j_3)^\gamma \beta(\ell,j_1,j_2,j_3)^s}{(1+S(\ell,j_1,j_2,j_3))^2} |z_{j_1}||z_{j_2}||z_{j_3}|\right)^2.
\end{multline*}
We have to decompose the last sum depending on whether $|\ell| \leq |j_2|$ or not. 

\textbf{First case} $|\ell| \leq  |j_2|$ 

In that case we can write
\begin{equation}
\label{pouetpouet}
 \sum_{\ell \in \Zc} |\ell|^{-2s}\left(\sum_{|j_1|\geq|j_2|\geq|j_3|,|\ell|} \frac{\mu(\ell,j_1,j_2,j_3)^\gamma \beta(\ell,j_1,j_2,j_3)^s}{(1+S(\ell,j_1,j_2,j_3))^2} |z_{j_1}||z_{j_2}||z_{j_3}|\right)^2.
\end{equation}
For a fixed $\ell$, the sum in $j_3$ can be bounded by 
$$
\sum_{|j_3| \geq \ell}|j_3|^\gamma |z_3| + \sum_{|j_3| \leq \ell}\ell^\gamma |z_3| \leq C \ell^\gamma\Norm{z}{s}
$$
using \eqref{peou}, provided $s - \gamma > 1/2$. Hence the expression \eqref{pouetpouet} is bounded by 
$$ 
C \Norm{z}{s}^2 \sum_{\ell \in \Zc} |\ell|^{-2(s-\gamma)}\left(\sum_{|j_1|\geq|j_2|} \frac{ |j_1|^s|z_{j_1}||j_2|^s |z_{j_2}|}{(1+|j_1|-|j_2|)^2}\right)^2.
$$
As the sequence $b=(|j|^s|z_{j}|)_{j\in \Zc}$ belongs to $\ell^2(\Zc)$ and the sequence $a=((1+|j|)^{-2})_{j\in \Zc}$ belongs to $\ell^1(\Zc)$, the convolution $a\star b$ belongs to $\ell^2(\Zc)$ and $\Norm{a\star b}{2}\leq C\Norm{z}{s}$. Therefore 
$$|\sum_j b_j (a\star b)_j|\leq C \Norm{z}{s}^2$$
which leads to 
$$
 \sum_{\ell \in \Zc} |\ell|^{-2s}\left(\sum_{|j_1|\geq|j_2|\geq|j_3|,|\ell|} \frac{\mu(\ell,j_1,j_2,j_3)^\gamma \beta(\ell,j_1,j_2,j_3)^s}{(1+S(\ell,j_1,j_2,j_3))^2} |z_{j_1}||z_{j_2}||z_{j_3}|\right)^2\leq C\Norm{z}{s}^6
$$
as expected.

\textbf{Second  case} $|\ell| \geq  |j_2|$ 

In this case we can write
\begin{align*}
 \sum_{\ell \in \Zc} |\ell|^{-2s}&\left(\sum_{|j_1|,|\ell|\geq|j_2|\geq|j_3|} \frac{\mu(\ell,j_1,j_2,j_3)^\gamma \beta(\ell,j_1,j_2,j_3)^s}{(1+S(\ell,j_1,j_2,j_3))^2} |z_{j_1}||z_{j_2}||z_{j_3}|\right)^2\\
\leq  &C \Norm{z}{s}^4 \sum_{\ell \in \Zc} |\ell|^{-2s}\left(\sum_{|j_1|} \frac{ |j_1|^s\ell^s}{(1+||j_1|-|\ell ||)^2} |z_{j_1}|\right)^2\\
\leq & C \Norm{z}{s}^4 \sum_{\ell \in \Zc} |(a\star b)_\ell |^2\leq C \Norm{z}{s}^6
\end{align*}
where we used \eqref{peou} and the notations introduced in the previous case.

\end{Proofof}

\medskip

\begin{Proofof}{Lemma \ref{lem:2}}

Let $G(z)=\sum_{\jb \in \Zc^n}b_\jb z_\jb$ be in $\Gamma^\gamma_n$ and $F=(F_j)_{j\in \Zc}$ a homogeneous vector field in $\Tc^{\infty,\nu}_m$ with
$$F_j(z) =  \sum_{\lb \in \Zc^m} a_{j\lb} z_{j\lb}, \quad j\in \Zc.$$
One has
\begin{align*}
\Lc_F G(z)&= \sum_k F_k(z) \frac{\partial G}{\partial z_k}(z)\\
&=  \sum_{k\in \Zc} \sum_{\lb \in \Zc^m}\sum_{\jb \in \Zc^n} a_{k\lb}z_\lb \sum_{i=1}^{n-1}b_{j_1\ldots j_ikj_{i+1}\ldots j_n}z_\jb.
\end{align*}
So in view of the symmetry in the estimates of the coefficients $a$ or $b$, one has to prove that,  there exist an integer $N$  and a constant $C>0$ such that, uniformly with respect to $\lb \in \Zc^m$ and   $\jb \in \Zc^n$, one has
\begin{equation}
\label{groumph}
\sum_{k\in \Zc} \frac{\mu(k,\lb)^{N+\nu}}{(\mu(k,\lb)+S(k,\lb))^N}
\mu(k,\jb)^\gamma \frac{\beta(k,\jb)^s}{(1+S(k,\jb))^2}
\leq C \mu(\lb,\jb)^\alpha \frac{\beta(\lb,\jb)^s}{(1+S(\lb,\jb))^2}.
\end{equation}

To prove this relation, we will show that there exist an integer $M$  and  a constant $C>0$ such that, uniformly with respect to $k \in \Zc$, $\lb \in \Zc^m$ and   $\jb \in \Zc^n$, the following relation holds: 
\begin{equation}\label{siduele}
\left[ \frac{\mu(k,\lb)}{\mu(k,\lb)+S(k,\lb)}\right] ^M\mu(k,\lb)^{\nu +4}
\mu(k,\jb)^\gamma{\beta(k,\jb)^s}
\leq C \mu(\lb,\jb)^\alpha {\beta(\lb,\jb)^s}.
\end{equation}
Indeed, if this relation is satisfied, then taking $N = M+4$, the relation \eqref{groumph} reduces to 
$$
\sum_{k \in \Zc} \frac{1}{(\mu(k,\lb)+S(k,\lb))^4( 1+ S(k,\jb))^2}
\leq \frac{C}{(1 + S(\lb,\jb))^2}. 
$$
Now as $\mu(k,\lb) \geq 1$ and as  for any $\lb \in \Zc^m$ and   $\jb \in \Zc^n$, one has
$$(1+S(k,\lb))(1+S(k,\jb))\geq (1+S(\lb,\jb))$$
we conclude using the fact that 
$$
\sum_{k\in \Zc}(1+S(k,\lb))^{-2}\leq C$$
where the constant is independent of $\lb \in  \Zc^m$.

\medskip
The rest of the proof consists in showing \eqref{siduele}. 

We assume without lost of generality that $\lb$ and $\jb$ are ordered (i.e.$|\ell_1|\geq |\ell_2|\geq \ldots \geq |\ell_m|$ and  $|j_1|\geq |j_2|\geq \ldots \geq |j_n|$ 
and we consider three different cases:

\medskip

\textbf{First case :} $\mu(k,\lb)\leq \mu(\lb,\jb)$ and $ \mu(k,\jb)\leq \mu(\lb,\jb)$.

\medskip

In this case it remains to prove (choosing $\alpha=\nu+\gamma+4$) that there exist $M$ and $C$ such that 
uniformly with respect to $\lb \in \Zc^m$ and   $\jb \in \Zc^n$
\begin{equation}\label{siduelemenos}
\left[ \frac{\mu(k,\lb)}{\mu(k,\lb)+S(k,\lb)}\right] ^M{\beta(k,\jb)^s}
\leq C {\beta(\lb,\jb)^s}.
\end{equation}
This is trivially true (with $M=0$ and $C=1$) if $|j_2|\geq |k|$ since then ${\beta(k,\jb)}
\leq  {\beta(\lb,\jb)}$. Now, if $|j_2|\leq |k|$, then ${\beta(k,\jb)}=|k||j_1|$ and 
\begin{itemize}
\item  \textbf{either} $S(k,\lb)\leq |k|/2$ and in that case $|\ell_1|\geq |k|/2$ and thus $\beta(\lb,\jb)\geq |\ell_1||j_1|\geq \frac 1 2 \beta(k,\jb)$ and \eqref{siduelemenos} is satisfied with $M=0$ and $C=2^s$.
\item \textbf{or} $S(k,\lb)\geq |k|/2$ and in that case
\begin{align*}
\frac{\mu(k,\lb)}{\mu(k,\lb)+S(k,\lb)}\beta(k,\jb)&\leq \frac{\mu(k,\lb)}{1+|k|/2}|k||j_1|\\
&\leq |\ell_1| |j_1|\frac{|k|} {1+|k|/2}\leq 2\beta(\lb,\jb)
\end{align*}
and \eqref{siduelemenos} is satisfied with $M=s$ and $C=2^s$.
\end{itemize}
\medskip

\textbf{Second case :} $\mu(k,\lb)> \mu(\lb,\jb)$.

\medskip

In this case,  $\mu(k,\lb)=\min(|\ell_2|,|k|)$ and $ \mu(\lb,\jb)\geq \min(|\ell_2|,|j_1|)$ and  therefore $\min(|k|, |\ell_2|)\geq |j_1|$. This in turn implies
$$\mu(k,\lb)\leq |\ell_2|,\quad \beta(k,\jb)=|j_1||k| \quad \mbox{ and }\quad \mu(k,\jb)\leq \mu(\lb,\jb)$$
and in the other hand
$$\mu(\lb,\jb)\geq |j_1|  \quad \mbox{ and }\quad \beta(\lb,\jb)=|\ell_1||\ell_2|.$$
Thus
$$\mu(\lb,\jb)\beta(\lb,\jb)\geq | j_1 ||\ell_1||\ell_2| \quad \mbox{ and }\quad \mu(k,\lb)\beta(k,\jb)\leq |\ell_2||j_1||k| $$
and 

\begin{itemize}
\item \textbf{either} $S(k,\lb)\leq |k|/2$ and in that case $|\ell_1|\geq |k|/2$ and thus 
$$\mu(\lb,\jb)\beta(\lb,\jb)\geq \frac 1 2 \mu(k,\lb)\beta(k,\jb) \quad \mbox{ and }\quad \beta(\lb,\jb)\geq  \frac 1 2 \beta(k,\jb)$$
 and \eqref{siduele} is satisfied with $M=0$, $\alpha=\nu+\gamma+2$ and $C=2^s$(here we use $s\geq \nu +4$).\\
 \item  \textbf{or} $S(k,\lb)\geq |k|/2$ and in that case we still have as in the first case
$$\frac{\mu(k,\lb)}{\mu(k,\lb)+S(k,\lb)}\beta(k,\jb)\leq 2\beta(\lb,\jb)$$
but furthermore
\begin{align*}
\frac{\mu(k,\lb)}{\mu(k,\lb)+S(k,\lb)}\mu(k,\lb)\beta(k,\jb)&\leq \frac{|\ell_2|^2|k||j_1|}{1+|k|/2}\\
&\leq |\ell_2||\ell_1|  |j_1|\frac{|k|} {1+|k|/2}\leq 2\beta(\lb,\jb)\mu(\lb,\jb)
\end{align*}
and \eqref{siduele} is satisfied with $M=s$, $\alpha=\nu+\gamma+2$ and $C=2^s$ (here we use again that $s\geq \nu +4$) .
\end{itemize}

\medskip

\textbf{Third case :} $\mu(k,\jb)> \mu(\lb,\jb)$.

As in the second case, $\mu(k,\jb)> \mu(\lb,\jb)$ implies $\min(|k|, |j_2|)\geq |\ell_1|$. This in turn implies
$$\mu(k,\jb)=\min(|k|, |j_2|),\quad \beta(k,\jb)=|j_1|\max(|k|,|j_2|) \quad \mbox{ and }\quad \mu(k,\lb)\leq \mu(\lb,\jb)$$
and in the other hand
$$\mu(\lb,\jb)\geq |\ell_1|  \quad \mbox{ and }\quad \beta(\lb,\jb)=|j_1||j_2|.$$
Thus
$$\mu(\lb,\jb)\beta(\lb,\jb)\geq | \ell_1 ||j_1||j_2| \quad \mbox{ and }\quad \mu(k,\jb)\beta(k,\jb)\leq |j_2||j_1||k|$$
and 
\begin{itemize}
\item  \textbf{either} $S(k,\lb)\leq |k|/2$ and in that case $|\ell_1|\geq |k|/2$ and thus 
$$\mu(\lb,\jb)\beta(\lb,\jb)\geq \frac 1 2 \mu(k,\jb)\beta(k,\jb) \quad \mbox{ and }\quad \beta(\lb,\jb)\geq  \frac 1 2 \beta(k,\jb)$$
 and \eqref{siduele} is satisfied with $M=0$, $\alpha=\nu+\gamma+2$ and $C=2^s$(here we use $s\geq \gamma $).\\
\item \textbf{or} $S(k,\lb)\geq k/2$ and in that case we still have as in the first case
$$\frac{\mu(k,\lb)}{\mu(k,\lb)+S(k,\lb)}\beta(k,\jb)\leq 2\beta(\lb,\jb)$$
but furthermore
\begin{align*}
\frac{\mu(k,\lb)}{\mu(k,\lb)+S(k,\lb)}\mu(k,\lb)\beta(k,\jb)&\leq \frac{|\ell_2||j_2||k||j_1|}{1+|k|/2}\\
&\leq |\ell_1||j_2|  |j_1|\frac{|k|} {1+|k|/2}\leq 2\beta(\lb,\jb)\mu(\lb,\jb)
\end{align*}
and \eqref{siduele} is satisfied with $M=s$, $\alpha=\nu+\gamma+2$ and $C=2^s$ (here we use again that $s\geq \gamma$) .

\end{itemize}

\end{Proofof}

\end{document}